\documentclass[aps,prl,showpacs,twocolumn]{revtex4-2}
\usepackage{graphicx}
\usepackage{subfig}
\usepackage{tabularx}
\usepackage{bm}
\usepackage{amsmath}
\usepackage{esvect}
\usepackage{mathtools}
\usepackage{xcolor}
\providecommand{\abs}[1]{\left\lvert#1\right\rvert}

\providecommand{\bra}[1]{\langle #1 \rvert}
\providecommand{\ket}[1]{\lvert #1 \rangle}

\providecommand{\be}{\begin{equation}}
\providecommand{\ee}{\end{equation}}
\providecommand{\ba}{\begin{eqnarray}}
\providecommand{\ea}{\end{eqnarray}}

\usepackage{mathbbol}

\usepackage{tikz}
\usepackage{amsfonts,amssymb}
\usepackage{dsfont}
\usepackage{physics}
\usepackage{tocvsec2}

\usepackage{graphicx}
\usepackage{subfig}
\usepackage{tabularx}
\usepackage{bm}
\usepackage{amsmath}
\usepackage{esvect}

\usepackage{siunitx}
\usepackage{hyperref}
\usepackage[capitalize]{cleveref}

\providecommand{\abs}[1]{\left\lvert#1\right\rvert}

\providecommand{\bra}[1]{\langle #1 \rvert}
\providecommand{\ket}[1]{\lvert #1 \rangle}

\providecommand{\be}{\begin{equation}}
\providecommand{\ee}{\end{equation}}
\providecommand{\ba}{\begin{eqnarray}}
\providecommand{\ea}{\end{eqnarray}}
\usepackage{amsfonts,amssymb}
\usepackage{dsfont}
\usepackage{physics}
\usepackage{tocvsec2}
 \DeclareUnicodeCharacter{202F}{\,}
\hypersetup{colorlinks=true,linkcolor=blue,citecolor = blue, urlcolor=black}
\newcommand{\beq}{\begin{equation}}
\newcommand{\eeq}{\end{equation}}

\usepackage{appendix}

\usepackage{graphicx}
\usepackage{subfig}
\usepackage{tabularx}
\usepackage{bm}
\usepackage{amsmath}

\providecommand{\abs}[1]{\left\lvert#1\right\rvert}

\providecommand{\bra}[1]{\langle #1 \rvert}
\providecommand{\ket}[1]{\lvert #1 \rangle}

\providecommand{\be}{\begin{equation}}
\providecommand{\ee}{\end{equation}}
\providecommand{\ba}{\begin{eqnarray}}
\providecommand{\ea}{\end{eqnarray}}

\usepackage{mathbbol}

\usepackage{tikz}
\usepackage{amsfonts,amssymb}
\usepackage{amsthm}
\usepackage{graphicx}
\usepackage{tabularx}
\usepackage{bm}
\usepackage{amsmath}
\usepackage{hyperref}
\usepackage[capitalize]{cleveref}

\providecommand{\abs}[1]{\left\lvert#1\right\rvert}

\providecommand{\bra}[1]{\langle #1 \rvert}
\providecommand{\ket}[1]{\lvert #1 \rangle}

\providecommand{\be}{\begin{equation}}
\providecommand{\ee}{\end{equation}}
\providecommand{\ba}{\begin{eqnarray}}
\providecommand{\ea}{\end{eqnarray}}
\usepackage{amsfonts,amssymb}
 \DeclareUnicodeCharacter{202F}{\,}
\hypersetup{colorlinks=true,linkcolor=blue,citecolor = blue, urlcolor=black}

\usepackage{appendix}

\begin{document}

\title{Non-Gaussianity from superselection rules}

\author{Nicolas Moulonguet$^{1,2}$}
\author{Eloi Descamps$^{1}$}
\author{José Lorgeré$^3$}
\author{Astghik Saharyan$^{1}$}
\author{ Arne Keller$^{1,4}$}
\author{Pérola Milman$^{1}$ }
\email{corresponding author: perola.milman@u-paris.fr}

\affiliation{$^{1}$Université Paris Cité, CNRS, Laboratoire Matériaux et Phénomènes Quantiques, 75013 Paris, France}
\affiliation{$^{2}$ Département de Physique de l’Ecole Normale Supérieure - PSL, 45 rue d’Ulm, 75230, Paris Cedex 05, France}
\affiliation{$^{3}$ Département de Mathématiques et Applications - PSL, 45 rue d’Ulm, 75230, Paris Cedex 05, France}
\affiliation{$^{4}$ Département de Physique, Université Paris-Saclay, 91405 Orsay Cedex, France}
\begin{abstract}
The quantum theory of the electromagnetic field enables the description of multiphoton states exhibiting nonclassical statistical properties, often reflected in non-Gaussian phase-space distributions. While non-Gaussianity alone does not fully characterize quantum states, several classifications have been proposed to hierarchize non-Gaussian states according to physically or informationally relevant resources. Here, we provide a physical interpretation of non-Gaussianity and connect it to a computational perspective by showing how a prominent classification—the stellar rank—emerges as a limiting case of the roots of polynomials that univocally represent bosonic states defined with a quantized phase reference, namely the Majorana polynomials. A direct consequence of our results is a revised interpretation of both the stellar rank and non-Gaussianity itself: when superselection rules are properly taken into account, quadrature non-Gaussianity - and nonzero stellar rank - act as witnesses of particle entanglement, rather than being linked with photon addition to Gaussian states as previously assumed. In addition, we show that because the stellar rank depends on a specific choice of coherent states, its relation to computational resources and potential quantum advantage is inherently basis-dependent, being naturally tied to quadrature eigenstates as the computational basis. Motivated by this observation, we generalize the notion of stellar rank to arbitrary computational bases, thereby establishing it as a genuine witness of bosonic resources that may enable quantum advantage.
\end{abstract}

\maketitle

{\it Introduction.}—Pure quantum states of the electromagnetic field are commonly described as coherent superpositions of photon-number eigenstates, giving rise to measurable interferometric and statistical signatures that have no classical counterpart. Phase-space quasiprobability distributions, such as the Wigner function \cite{PhysRev.40.749}, provide a widely used framework aiming to characterize these nonclassical properties. Within this description, a single bosonic mode is represented by two conjugate quadratures formally analogous to the position and momentum of a quantum harmonic oscillator.

In quadrature phase space, coherent and squeezed states are represented by positive Gaussian distributions. Although these states can exhibit markedly different performances in certain quantum-information tasks \cite{RevModPhys.84.621}, including quantum metrology \cite{PhysRevA.98.012114,Maccone2020squeezingmetrology}, quantum protocols restricted to states, operations, and measurements with positive Wigner functions can be efficiently classically simulated \cite{PhysRevLett.109.230503,VeitchEtAl2012,Veitch2014,Bartlett2002,Bartlett2002Efficient,Gross2006Hudsons, Gross, PhysRevA.73.012301}. As a consequence, non-Gaussianity—corresponding, for pure states, to negativity of the Wigner function \cite{HUDSON1974249,Soto:1983avg,Gross2006Hudsons}—is identified as a necessary resource for potential quantum advantage in bosonic quantum information. Prominent examples of non-Gaussian states used to encode quantum information include single-photon states \cite{KLM,RevModPhys.79.135,PhysRevLett.39.691}, Schrödinger cat states \cite{Gilchrist_2004,PhysRevLett.119.030502,PRXQuantum.5.020355}, and Gottesman–Kitaev–Preskill (GKP) states \cite{gottesman_encoding_2001}. However, non-Gaussianity alone, does not guarantee computational advantage: several protocols involving highly non-Gaussian states remain efficiently classically simulatable \cite{Calcluth2022efficientsimulation,PhysRevResearch.3.033018,xmtw-g54f,PhysRevA.110.042402}. These observations motivate the search for refined classifications of non-Gaussian states and operations, or more generally of nonclassical features of the electromagnetic field, connecting them to physically or computationally meaningful criteria \cite{Hahn2025classicalsimulation,PhysRevLett.129.013602,PhysRevLett.123.043601,descamps2024superselectionrules,PhysRevLett.130.090602,PhysRevLett.99.250501}.

One such classification is provided by the stellar rank \cite{Stellar}. Any pure bosonic state expressed in the quadrature representation can be associated with a function whose zeros define a constellation in phase space. The number $r^{\star}$
 of such zeros, known as the stellar rank, defines a hierarchy of non-Gaussian states. Despite its usefulness, however, the physical interpretation of stellar rank and its relation to fundamental quantum resources remain only partially understood \cite{PhysRevLett.130.090602, provaznik2026,Fiurasek:22,PRXQuantum.2.020333}.

In this Letter, we show that stellar rank and quadrature non-Gaussianity in bosonic systems originate from particle entanglement when the phase reference is treated as a quantum degree of freedom. Within this framework, bosonic states admit a complete representation in terms of Majorana polynomials associated with a finite-dimensional symmetric Hilbert space that explicitly includes a quantum reference mode. This representation allows one to track and physically interpret the transition from finite-degree Majorana polynomials to normalizable functions in the Bargmann–Fock space \cite{Bargmann1961,Segal1963Relativistic}, which describes the continuous-variable (CV) limit \cite{PhysRevA.63.012102,Ricci1986}. In this limit, commonly used to represent bosonic states, the phase reference becomes implicit and effectively classical, and states can be described as single-mode states in the Hilbert space $\mathcal{H}_{\infty}=L^2(\mathbb{R})$. Our analysis of this limiting process in terms of bosonic resources shows that the normalization condition imposes nontrivial constraints: only states arising as limits of energy-constrained Majorana polynomials correspond to physical CV states. Consequently, the stellar rank acquires a clear physical meaning as a sufficient condition for particle entanglement, while remaining effectively bounded and much smaller than the total number of Majorana roots for all physical CV states. Finally, we analyze how the relation between stellar rank and computational resources depends on the choice of computational basis \cite{descamps2025HW}, and we introduce a generalized notion of stellar rank applicable to arbitrary encodings, providing a basis-dependent witness of potential quantum advantage beyond the quadrature setting.

{\it Methods:} In the CV formalism, single mode (in mode $a$) pure states of the field are represented as infinite superpositions of Fock states, $\ket{\psi}_{C}= \sum_{k=0}^{\infty} c_k \ket{k}_a \in {\cal H}_{\infty}$, with $\sum_{k=0}^{\infty}|c_k|^2=1$ and $\ket{k}=\frac{(\hat a^{\dagger})^k}{\sqrt{k!}}\ket{\emptyset}$. In \cite{Stellar}, the alternative representation is introduced:
\be\label{stellar}
 \ket{\psi}_{C} = {\mathcal{N}}\prod_{j=1}^{r^\star(\psi)}\hat D(z_j)\hat{a}^\dagger \hat D^\dagger(z_j) \ket{G_\psi},
 \ee
where ${\cal N}$ is a normalization constant, $\hat D(z_j)$ is the quadrature displacement operator, $z_j = e^{i\phi_j}\tan{\frac{{\theta_j}}{2}} \in \mathbb{C}$, $j \in \{1,...,r^\star(\psi)\}$, are the associated coherent state's amplitudes, $\ket{G_\psi}$ is a Gaussian state, and $r^\star(\psi)$ is the stellar rank. The stellar rank is the number of roots $z_j$  -counted with multiplicity - of the Bargmann function \cite{Bargmann1961} $B(z)=e^{\frac{|z|^2}{2}}\bra{z^*}\psi\rangle = \sum_{k=0}^\infty \frac{c_k}{\sqrt{k!}}z^k$, normalized with respect to the Gaussian measure, $\frac{1}{\pi}\int_{\mathbb{C}} |B(z)|^2 e^{-|z|^2} \, d^2 z = 1$. 
Zeros of $B(z)$ correspond to negativities of the Wigner function, and non-Gaussian states can be classified according to the stellar hierarchy, formally defined in \cite{Stellar}.  
Inspection of \eqref{stellar} provides an intuitive interpretation of  the stellar rank as the number of displaced photons added to a Gaussian state \cite{PhysRevLett.123.043601, Fiurasek:22}.  

\begin{figure}[h]  
    \centering
    \includegraphics[width=1.\columnwidth]{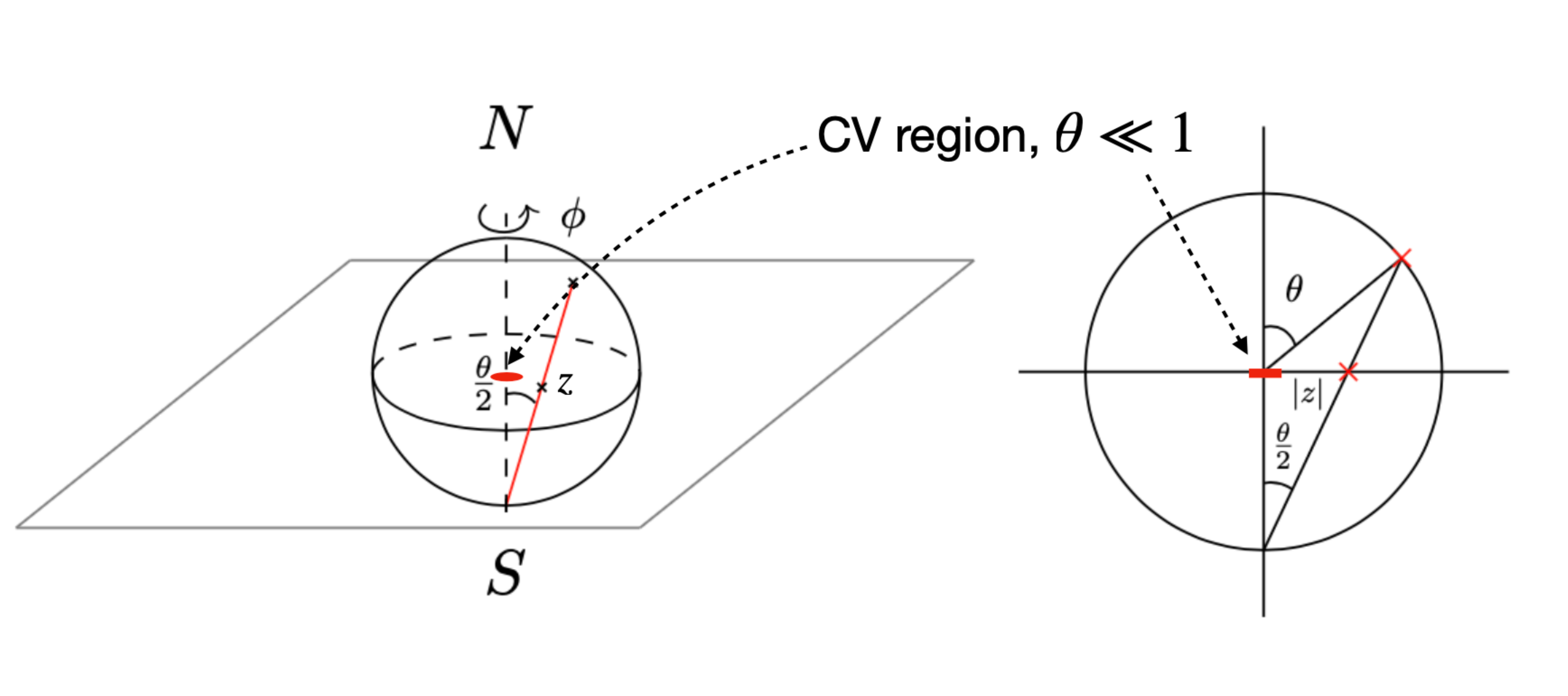}  
    \caption{
        (Color online) Principles of the inverse stereographic representation of the roots of the Majorana polynomial on the sphere: A root located at position $z$ in the complex plane is mapped onto the sphere by drawing a line from the south pole $S$ through $z$; the intersection of this line with the sphere defines its spherical position (left panel and cross-sectional view on the right). For example, the root $z=0$, corresponding to the state $\ket{N}_a$, is mapped to the north pole $N$. Roots approaching $S$ correspond to $\abs{z}\to\infty$ in the complex plane. The CV region (red, central region of the sphere and corresponding disk in the plane) is defined by $\abs{z}^2 \ll \sqrt{N}$, equivalently $\theta \ll 1$. In the CV limit, the spherical measure reduces to the planar Gaussian measure (see main text).
    }
    \label{fig1}
\end{figure}

As pointed out in  \cite{PhysRevA.68.042329, PhysRevA.55.3195, PhysRevA.58.4244,PhysRevA.58.4247,doi:10.1142/S0219749906001591,Sanders_2012,PhysRev.155.1428,RevModPhys.79.555}, within the CV formalism the phase reference is implicit and considered as classical. The CV representation can be recovered as a limiting case of a more general description of bosonic states in which a phase-reference mode is included explicitly and treated quantum mechanically, while respecting the photon-number superselection rule (SSR). In this framework, a minimal pure SSR-compliant (SSRC) state can be written as

\begin{eqnarray}\label{state}
&\ket{\psi}_S=\sum_{n=0}^N& c_n\ket{n}_a\ket{N-n}_b= \\
&&{\cal M}\prod_{k=1}^N \hat R(\theta_k,\phi_k)\hat b^{\dagger}\hat R^{\dagger}(\theta_k,\phi_k)\ket{\emptyset}\nonumber,
\end{eqnarray}
where ${\cal M}$ is a normalization factor, $\hat R(\theta,\phi)=e^{i\hat J_z \phi}e^{i\hat J_y \theta}$, with $\hat J_x=\frac{1}{2}(\hat a^{\dagger}\hat b +\hat b^{\dagger}\hat a)$, $\hat J_y=\frac{1}{2i}(\hat a^{\dagger}\hat b -\hat b^{\dagger}\hat a)$ and $\hat J_z=\frac{1}{2}(\hat a^{\dagger}\hat a -\hat b^{\dagger}\hat b)$ \cite{Schwinger}. The angular coordinates $\theta$,$\phi$ are defined as in Fig. \ref{fig1} of End Matter and $\hat R(\theta_k,\phi_k)\hat b^{\dagger}\hat R^{\dagger}(\theta_k,\phi_k)=e^{i\phi_k}\sin{\frac{\theta_k}{2}}\hat a^{\dagger}+\cos{\frac{\theta_k}{2}}\hat b^{\dagger}$. The SSRC representation is exact, it is not a truncation of the CV space, and it presents the formal advantage of dealing with finite total photon number (that can be arbitrarily large), which helps solving relevant problems in quantum optics based quantum information tasks \cite{descamps2024superselectionrules, descamps2025HW, saharyan2025, descamps2025unifiedframeworkbosonicquantum}. 

State \eqref{state} also admits an analytical representation \cite{AVourdas_2004, vourdas_2006} through the Majorana polynomial \cite{Majorana1932AtomiOI} $P_N\left (z\right )=\left (1+|z|^2\right )^{N/2}{}_{z^*}\langle N |\psi\rangle_S=\sum_{n=0}^N \sqrt{\binom{N}{n}}c_n z^n$, with $z=e^{i\phi}\tan{\frac{\theta}{2}}$, where $\ket{N}_z=(e^{i\phi}\sin{\frac{\theta}{2}}\hat a^{\dagger}+\cos{\frac{\theta}{2}}\hat b^{\dagger} )^N/\sqrt{N!}\ket{\emptyset}$ is a $N$ photon Fock state, the analogous to a spin coherent state \cite{Chryssomalakos_2018, RHoltz_1974}. 

The CV limit corresponds, in \eqref{state}, to restricting to states for which $\langle \hat a^{\dagger}\hat a\rangle \ll \sqrt{N}$. In this regime, $\ket{\psi}_S$ are well approximated by $\ket{\psi}_C$, in the sense that their analytical representation coincide, {\it i.e.}, $P_N\to B$ (see Supplementary Material \cite{SM}). This raises several questions: Is the CV limit the only way to connect $P_N$ and $B$? What are the physical consequences and the interpretation of this limit? And most importantly, how and when do stellar rank and non-Gaussianity emerge—or vanish—in this transition?

{\it Results:}  We begin by comparing the CV limit of $P_N\!\left(\frac{z}{\sqrt{N}}\right)$ (the change of variables $z \to \frac{z}{\sqrt{N}}$ implements the natural scaling associated with the CV limit) and its roots with the Bargmann function $B(z)$ for two representative states: Fock states and Schr\"odinger cat--like states. For Fock states $\ket{N}_{\frac{w}{\sqrt{N}}}$, $P_N^{F}\!\left(\frac{z}{\sqrt{N}}\right) = \frac{\left(1+\frac{w z}{N}\right)^N}{\left(1+\frac{|w|^2}{N}\right)^\frac{N}{2}}$, which can be compared with $B^{F}(z) = e^{w z} e^{-\frac{|w|^2}{2}}$, the Bargmann function of the Glauber coherent state $\ket{w}$. When $|w|^2 \ll \sqrt{N}$ (the CV limit) one has $\ket{N}_{\frac{w}{\sqrt{N}}} \to \ket{w}$ \cite{descamps2025unifiedframeworkbosonicquantum, descamps2024superselectionrules, saharyan2025}. Consistently, we notice that for $w, z$ fixed, $\lim_{N \to \infty} P_N^{F}\!\left(\frac{z}{\sqrt{N}}\right) \to B^{F}(z)$, suggesting that recovering $B^{F}$ from $P_N^{F}$ imposes constraints both on the state coefficients $c_k$ and on the domain where the Majorana roots may appear: indeed, conditions $|w|^2,|z|^2 \ll \sqrt{N}$ define a restricted region of the complex plane (Fig.~\ref{fig1} in End Matter) for both the state coordinate $w$ and the analytical coordinate $z$. The Majorana roots $z_j$, defined by $P_N^{F}\!\left(\frac{z_j}{\sqrt{N}}\right) = 0$, are $N$-fold degenerate and satisfy $\frac{z_j}{\sqrt{N}} = -\frac{\sqrt{N}}{w}$ for $j = 1,\ldots,N$. Hence, when $|w|/\sqrt{N} \ll 1$, one has $|z_i|/\sqrt{N} \gg 1$, placing the roots outside the CV-limit region where $P_N^{F} \to B^{F}$: For Fock states, either the state lies within the CV-limit region while the roots do not, or vice versa, consistent with the usual interpretation that the zeros of $B^{F}(z)$ occur ``at infinity''. This observation motivates at least two questions: what controls the existence of the limit $ P_N^{F}\!\left(\frac{z}{\sqrt{N}}\right) \to B^{F}(z)$? And in what precise sense are the zeros of $B^{F}(z)$ located ``at infinity'', {\it i.e.}, relative to which physical quantity should this notion of infinity be understood?

Our second example consists of Schr\"odinger cat--like states 
$\ket{{\cal S}}_{\frac{w}{\sqrt{N}}} = {\cal N}_S \left( \ket{N}_{\frac{w}{\sqrt{N}}} - \ket{N}_{\frac{-w}{\sqrt{N}}} \right)$, 
where ${\cal N}_S$ is a normalization factor depending on $N$ and $w$. 
The associated Majorana polynomial reads 
$P_N^S\!\left(\frac{z}{\sqrt{N}}\right) = \frac{\left(1+\frac{w z}{N}\right)^N - \left(1-\frac{w z}{N}\right)^N}{\left[ 2 \left(1-\cos^N{\theta}\right) \left(1+\frac{|w|^2}{N}\right)^N \right]^{1/2}}$, 
with roots $z_{k,S} = \frac{iN}{w}\tan\!\left(\frac{k \pi}{N}\right)$ for $k \in \{1,\ldots,N\}$. In the CV limit $\abs{w}^2 \ll \sqrt{N}$, $\ket{{\cal S}}_{\frac{w}{\sqrt{N}}} \to \ket{{\cal C}}_w = {\cal N}_C \left( \ket{w} - \ket{-w} \right)$, corresponding to the odd CV Schr\"odinger cat state, where ${\cal N}_C$ is a normalization constant. 
The Bargmann function of $\ket{{\cal C}}_w$ is $B^S(z) = -2i \left( \frac{e^{-|w|^2}}{2(1-e^{-2|w|^2})} \right)^{1/2}\sin\!\left(i w z\right)$, 
with roots $z_{k,C} = \frac{i k \pi}{w}$ for $k \in \mathbb{Z}$. 
For fixed $w$ and $z$, one has $\lim_{N\to \infty} P_N^S\!\left(\frac{z}{\sqrt{N}}\right) = B^S(z)$. The Bargmann function $B^S(z)$ thus has an infinite number of roots, which is possible in the limit $N \to \infty$. 
However, in order for $z_{k,S} \to z_{k,C}$, one must impose $\left|\tan\!\left(\frac{k \pi}{N}\right)\right| = \frac{|z_{k,S}||w|}{N} \ll 1$, 
so that $N \tan\!\left(\frac{k \pi}{N}\right) \simeq k \pi$. 
This condition corresponds to $k \ll N$, revealing a hierarchy between the limits involved. 
In particular, although $k$ becomes effectively unbounded in the CV limit, one must still require $k \ll N \to \infty$ for all $k$ that label $z_{k,C}$ in order to recover the correspondence between the Majorana roots and those of the Bargmann function.

These examples illustrate how different limiting procedures and scaling relations can be identified when trying to describe the transition from $P_N$ to $B$. 
We now establish a consistent framework relating these quantities and clarify their connection to the CV limit.

For this we analyze the normalization condition of the Majorana polynomial, given by
\begin{equation}\label{I1}
\frac{N+1}{N\pi}\int_{\mathbb{C}} d^2 z \frac{\abs{P_N\left (\frac{z}{\sqrt{N}}\right )}^2}{\left (1+\frac{|z|^2}{N}\right )^{N+2}}=1,
\end{equation}
No approximations have been made so far, and \eqref{I1} is equivalent to $\abs{{}_S\langle \psi | \psi \rangle_S}^2=\sum_{k=0}^N \abs{c_k}^2 = 1$ using \eqref{state}. We recall that the completeness relation associated with the stereographic projection of the sphere into the complex plane is given by $\frac{N+1}{\pi}\int_{\mathbb{C}} \frac{d^2 z}{\left (1+|z|^2\right )^2}\ket{N}_{z}\!\prescript{}{z}{\bra{N}}=\mathbb{1}$. 

The integral \eqref{I1} can always be split as
\begin{equation}\label{I2}
I_D+\varepsilon_D= \frac{1}{\pi} \int_{D} d^2 z \abs{P_N \left (\frac{z}{\sqrt{N}}\right )}^2 e^{-\abs{z}^2}+\varepsilon_D=1,
\end{equation}
where $D$ is a disk of radius $R$ centered at $0$ in the complex plane and $0 \leq \varepsilon_D \leq 1$. Since the Bargmann function is normalized on $\mathbb{C}$ with respect to the Gaussian measure, we seek conditions on $P_N$ (and possibly on the distribution of its zeros) and on the domain $D$ such that $\varepsilon_D$ is as small as one wishes, and examine whether this implies $P_N \to B$. Importantly, we do not impose from the outset the CV limit — previously identified as the regime $|z|^2 \le R \ll \sqrt{N}$. Rather, we ask whether $I_D \to 1$ can occur in more general domains and, in particular, whether the CV limit is not only sufficient but also necessary to establish the correspondence between the Bargmann function and the Majorana polynomial.

If the CV limit is assumed, namely $z \in D \Rightarrow |z|^2 \ll \sqrt{N}$ with $N \gg 1$, then $\frac{N+1}{N\left(1+\frac{|z|^2}{N}\right)^{N+2}} \to e^{-|z|^2}$ in \eqref{I1}. In this regime, the measure converges to the Gaussian one, so $I_D \to 1$ provided the normalization mass is sufficiently localized within $D$ (see Supplementary Material \cite{SM}). We therefore ask under which conditions such localization occurs, and whether the CV limit is also necessary for having $\varepsilon_D \to 0$ in \eqref{I2}. We first examine whether a general polynomial $P_N$ can be normalized on $\mathbb{C}$ with respect to the Gaussian measure. Taking $D \to \mathbb{C}$ yields $I_{\mathbb{C}} = \sum_{j=0}^N \binom{N}{j} |c_j|^2 \frac{j!}{N^j} \leq 1$ for $N>1$, showing that arbitrary $P_N$, \textit{i.e.} without constraints on the coefficients $c_j$, cannot be normalized using \eqref{I2}. This reflects the stronger suppression of high-degree contributions by the Gaussian measure, when compared with the Majorana polynomial measure in Eq. \eqref{I1}. Achieving $I_D \to 1$ therefore requires a specific structure of $P_N$.

When the radial growth is dominated by degree $k$, the leading contribution behaves as $|c_k|^2 |z|^{2k} e^{-|z|^2}$, with maximum at $|z_m|^2 = k$. The Gaussian weight is thus concentrated within $|z|\lesssim \sqrt{k}$. For $P_N$ dominated by degree $N$, the integrand peaks at $|z|\sim \sqrt{N}$; a domain of this scale captures the full mass of any polynomial $P_N$ but does not ensure normalization. We therefore restrict to a bounded domain $D=\{|z|\le R\}$ with $R<\sqrt{N}$ and consider polynomials whose growth in $D$ is effectively controlled by degrees $K < N$. We assume that for any fixed $D$ and $\eta>0$, there exists a degree $K \ll \sqrt{N}$ such that for all $|z|\le R$, $\sum_{n>K}^{N}\sqrt{\binom{N}{n}}\,|c_n|\,\frac{|z|^n}{\sqrt{N^n}} \le \eta$, so that $P_N$ is uniformly approximated by its truncation up to order $K$, $P_{N,K}$.

Since $K \ll \sqrt{N}$, Stirling’s approximation gives $P_N\!\left(\frac{z}{\sqrt{N}}\right ) \simeq  P_{N,K}\!\left(\frac{z}{\sqrt{N}}\right) \simeq \sum_{n=0}^K \frac{c_n}{\sqrt{n!}} z^n$ inside $D$, corresponding to a truncated Bargmann–Fock expansion \cite{Bargmann1961, Segal1963Relativistic}. For $z\in D$, $I_D = \sum_{n=0}^K |c_n|^2 \frac{\int_0^R |z|^{2n} e^{-|z|^2}\,d(|z|^2)}{n!}$. We see that choosing $K \ll R^2$ (see Supplementary Material \cite{SM}), the integral approaches $n!$ for all $n \le K$, and $I_D \to 1$ up to exponentially small corrections, showing that normalization is asymptotically possible. We have shown that the existence of $P_{N,K}$ as previously defined implies normalization. We show in the Supplementary Material \cite{SM} the reciprocal, {\it i.e.}, that normalization in $\mathbb{C}$ implies the existence of $P_{N,K}$ as previously defined. In these cases, the normalization mass is localized within $|z|\lesssim \sqrt{K} \ll N^{1/4}$, which is precisely the CV limit, and the state is normalized.

We now connect these results with the Bargmann function and the stellar rank. Let $D\subset\mathbb{C}$ be a compact disk containing the region where the normalization mass is concentrated. Because the coefficients of $P_{N,K}$ are bounded and the truncations are uniform on compact sets, the truncated polynomials $P_{N,K}$ converge uniformly on $D$ to a holomorphic limit function $\mathcal{B}$ by Montel's theorem \cite{Conway}. By Hurwitz’s theorem  \cite{Conway}, the zeros of $\mathcal{B}$ in $D$ are limits of zeros of $P_{N,K}$. Moreover, the coefficients of $P_{N,K}$ obtained using the Stirling approximation coincide with the truncated Bargmann–Fock expansion of a normalized state, as mentioned. Consequently, the limit function $\mathcal{B}$ coincides with the Bargmann function $B$, which is entire and normalized in Bargmann–Fock space. The zeros of $B$ therefore arise as limits of the zeros of the truncated polynomial $P_{N,K}$ in an arbitrarily large domain $D$, providing the stellar rank.

{\it Discussion:} A first consequence of the present physically motivated approach is that the CV limit is both necessary and sufficient for representing bosonic states in the form \eqref{state} as normalizable functions in the Bargmann–Fock space.

A second important result is that, although the stellar rank $r^{\star}$ may be formally unbounded, it is not possible to associate all $N$ roots of a Majorana polynomial with the stellar rank. Instead, one finds the scaling bound $r^{\star} \le K \ll \sqrt{N}$. Thus only a subset of roots contributes in the CV limit, which reflects the fact that the CV description is inherently incomplete and clarifies why the stellar rank does not provide full information about the state, but only about its non-Gaussian features. The obtained scaling of $r^{\star}$ is consistent with (but more restrictive than) the Schrödinger-cat example, where the weaker condition $K \ll N$ was empirically identified as necessary to extract the stellar rank from the Majorana roots in the CV limit.

Third, a detailed analysis of the transition from the Majorana polynomial to the Bargmann function clarifies the relation between finite-dimensional states and their CV description \cite{Arzani2025BosonicComplete,upreti2025, Marshall:23, orlov2025, maltesson2025}. States in ${\cal H}_{\infty}$ arise as mathematical limits of truncated physical states that are normalized on finite domains. In this sense, it is more accurate to say that the Bargmann function $B$, which represents states in ${\cal H}_{\infty}$, can approximate finite-dimensional physical states arbitrarily well, rather than the converse, as is often implicitly assumed in the literature. Due to this limit, some states, as Schrödinger cat-like ones, can be associated with an infinite stellar rank. However, physical pure states are fundamentally defined in a finite-dimensional Hilbert space ${\cal H}_{N+1}$ associated with the SSRC representation \eqref{state}. The CV limit does not encompass all such states; instead, it selects those that can be approximated arbitrarily well by states supported in reduced subspaces ${\cal H}_{K+1}$ with $K \ll \sqrt{N}$. Only this restricted family admits a consistent CV representation while retaining part of the physically relevant structure encoded in the Majorana roots, which, in the limit, become the stellar rank.

Finally, we examine the connection with particle entanglement. A symmetric $N$-boson state $\ket{\psi}_S$ is particle-separable iff there exist angles $\theta,\phi$ such that
$\ket{\psi}_S=\hat R(\theta,\phi)\ket{N}_b \equiv \ket{N}_w$, {\it i.e.}, iff it is a Fock state on an arbitrary mode $w$. Otherwise, the state is particle-entangled. In the CV limit, when $\sqrt{N}\gg |w|^2$, one has $\ket{N}_{w/\sqrt{N}} \to \ket{w}_a$, which is a coherent state and satisfies $r^{\star}(w)=0$. It follows that
$r^{\star}(\psi)\neq 0$  implies particle entanglement. The converse, however, does not hold: continuous-variable squeezed states are limits of particle-entangled SSRC states \cite{descamps2025unifiedframeworkbosonicquantum} while still satisfying $r^{\star}(\psi)=0$. Thus, a nonzero stellar rank only witnesses particle entanglement, in the sense that particle entanglement is necessary for $r^{*}\neq 0$ but not sufficient.

This connection between stellar rank and particle entanglement is not apparent in the standard CV formalism, where the phase reference is implicit and effectively classical. Making the reference explicit reveals that single-mode non-Gaussianity and non-Gaussian entanglement share a common physical origin \cite{MattiaX}. In particular, non-Gaussian CV states arise as limits of SSRC particle-entangled states that are intrinsically multimode (once the phase reference is included) and cannot be disentangled by passive transformations such as rotations $\hat R(\theta,\phi)$, requiring operations that are the analogous to spin-squeezing transformations \cite{PhysRevLett.86.4431, daltonNewSpinSqueezing2014a, daltonQuantumEntanglementSystems2017, PhysRevA.102.012412, PhysRevLett.132.153601}. Including the phase reference then refines the common interpretation of stellar rank as related to “single-photon addition”.

{\it Generalization:}  Using the established relation between the Bargmann function and the Majorana polynomials, we can exploit the results of \cite{descamps2025HW} to consistently connect the stellar rank with potential quantum advantage in an arbitrary computational basis. To illustrate this, we analyze an arbitrary SSRC state $\ket{\psi} = \hat U \ket{N}_b$. If the unitary $\hat U$ is not a rotation $\hat R(\theta, \phi)$, then $\hat U$ is a SSRC operation that necessarily generates particle entanglement \cite{PhysRevLett.132.153601, descamps2025unifiedframeworkbosonicquantum} and can, in the CV limit, be associated with $r^{\star}(\psi)\neq 0$, in which case, it serves as a resource for potential quantum advantage. The Majorana polynomial associated with $\ket{\psi}$ can be written as
$P_N\!\left(\frac{z}{\sqrt{N}}\right)
=\left(1+\frac{|z|^2}{N}\right)^{N/2}
{}_{\frac{z^*}{\sqrt{N}}}\!\bra{N}\hat U\ket{N}_b
= \tilde P_N^{F}\!\left(\frac{z}{\sqrt{N}}\right)$,
where $\tilde P_N^{F}$ denotes the Majorana polynomial of the Fock state $\ket{N}_b$ expressed in a unitarily transformed basis of ${\cal H}_{N+1}$, {\it i.e.} $\ket{N}_{\frac{z}{\sqrt{N}}} \mapsto \hat U \ket{N}_{\frac{z}{\sqrt{N}}}$.

In the appropriate CV limit \cite{descamps2025HW}, the mapping $\ket{N}_b \mapsto \hat U\ket{N}_b$ defines the vacuum of a new phase space that is unitarily equivalent to the usual quadrature phase space. Notice that such transformations of the phase reference are not accessible in all generality within the standard CV description. In addition, as shown in \cite{descamps2025HW}, such unitarily equivalent phase spaces correspond to different computational bases. In particular, Gaussian states and operations in the quadrature representation arise as the $d\to\infty$ limit of states and operations defined in a specific qudit phase space that are provably efficiently classically simulable \cite{Gross2006Hudsons,Veitch2014,VeitchEtAl2012,PhysRevA.73.012301, PhysRevA.71.042302}.

Changing to a unitarily equivalent phase-space representation therefore modifies both the notion of Gaussianity and the phase reference with respect to which the phase space is defined.  As a result, a Gaussian distribution in one phase space may appear non-Gaussian in another, thereby acquiring a nonzero stellar rank. In this case, similarly to quadrature non-Gaussian states, non-Gaussian states in different phase spaces are associated with potential quantum advantage, but now relative to the transformed computational basis. The stellar rank defined in the quadrature phase space therefore represents only a particular instance of a broader notion of non-Gaussianity: however, when generalized to unitarily equivalent phase spaces, it provides a computational basis-dependent witness of potential quantum advantage. This discussion reflects the fact that potential quantum advantage is not an intrinsic property of states and operations, but rather a relational property between different families of operations, as discussed in \cite{Wagner:2025fmv,descamps2025unifiedframeworkbosonicquantum,descamps2025HW}. One application of our results is the extension of the framework of Ref. \cite{PhysRevLett.130.090602}, for instance, beyond the quadrature representation to arbitrary computational bases.

{\it Conclusion:}  We establish an explicit connection between the stellar rank—currently used to classify non-Gaussian states—and the roots of the Majorana polynomial, which fully characterize bosonic states when the phase reference is treated as a quantum degree of freedom. Although it was known that the stellar rank emerges from Majorana roots in the $N \to \infty$ limit \cite{UlysseHolo}, a detailed understanding of this limit—including its mathematical subtleties, parameter scalings, and physical interpretation—has remained only partially understood. Previous approaches relied on analogies between atomic and optical systems \cite{RevModPhys.90.035005} or on particular families of states \cite{PhysRevA.6.2211,PhysRevA.68.033821}, and largely ignored the role of bosonic exchange symmetry and the associated superselection rules. As a result, the $N \to \infty$ limit has often been interpreted simply as a transition from a finite-dimensional qudit system to an unconstrained infinite-dimensional Fock space, thereby obscuring the rich underlying structure and physical constraints of this limit, which are essential for the definition of normalized CV states and for relevant nonclassical CV properties such as non-Gaussianity.

Our results instead show that particle entanglement provides the physical origin of stellar rank and quadrature non-Gaussianity, even in the single-mode CV regime. They can be {\it both} seen as a formal tool that permits a better understanding of quantum optical phenomena in the CV regime by showing that they can always be mapped into a finite-dimensional space and interpreted from a first-quantization-oriented perspective {\it and} as a way to better identify the fundamental physical mechanisms underlying quantum optical phenomena in concrete physical situations. In this latter case, the SSRC representation makes explicit the role of particle entanglement—an aspect that is not clearly accounted for within the CV formalism—as a key ingredient underlying non-Gaussianity. This perspective naturally leads to a consistent generalization of stellar rank to arbitrary phase-space representations associated with different computational bases.


More broadly, our work highlights the fundamental link between nonclassicality and particle entanglement in the quantum optical regime. These connections have been extensively studied in symmetric many-particle systems \cite{Giraud_2010,PhysRevLett.103.070503,PhysRevLett.106.180502,Albach,PhysRevA.87.012104,Aulbach2010MaxEntSymmetric,PhysRevA.105.022433,PhysRevA.96.032304,Rudzinski2024orthonormalbasesof,PhysRevA.90.050302,10.1088/1361-6633/ae440a,sanchezsoto2026quantumskymajoranastars}, and our results open the perspective of relating this literature to the CV limit. They support the view that a complete theory of CV quantum resources should incorporate this particle-based perspective and may naturally extend to other bosonic platforms, including Bose–Einstein condensates \cite{PhysRevA.85.040306,PhysRevA.88.023609}. Finally, a promising direction is the use of the present analysis for a better understanding of notions such as operational non-Gaussianity \cite{ldbf-vbyp}, with the aim of clarifying the formal and physical structures of relevant bosonic nonclassical resources.

{\it Acknowledgements:} We acknowledge funding from Plan France 2030 through Projects No. ANR-22-PETQ-0006 and No. ANR-24-CE97-0003 EQUIPPS. We also acknowledge Tim Heib for the careful reading of the manuscript.
\clearpage
\section{End Matter}

\subsection{The stereographic projection and the CV limit}

\onecolumngrid
\appendix

	\section{Some useful properties and definitions}
Before addressing the specific proofs discussed in the main text, we first remind some definitions and properties of Bargmann functions and Majorana polynomials.

Let's consider a CV monomode quantum state $\ket{\psi}$ and its expansion in Fock states $\ket{k}$: 
\[
\ket{\psi} = \sum_{k=0}^{\infty} c_k\ket{k};
\]
its associated Bargmann function $B_{\ket{\psi}}(z)$ is defined as:
\[
B_{\ket{\psi}}(z) = \sum_{k=0}^{\infty} c_k\frac{z^k}{\sqrt{k!}}.
\]
We recall that  the coefficients $c_k$ can be recovered from the Bargmann function as follow:
\beq
\label{eq:ckB}
c_k = \frac{1}{\pi}\int_{\mathbb{C}} B_{\ket{\psi}}(z) \frac{(z^*)^k}{\sqrt{k!}}e^{-\abs{z}^2}d^2z.
\eeq

The Bargmann function is normalized as
\beq
\label{eq:normBargmann}
\frac{1}{\pi} \int_{\mathbb{C}} 
\abs{B_{\ket{\psi}}(z)}^2 e^{-\abs{z}^2} d^2z = 
\sum_{k=0}^{\infty}\abs{c_k}^2 = 1.
\eeq

To the minimal $N$-photon SSRC pure state  $\ket{\psi}_s = \sum_{k=0}^{N}  c_k\ket{k}_a\ket{N-k}_b$, we can associate the Majorana polynomial:
\beq
\label{eq:majorPoly}
P_N(z) =  \sum_{k=0}^N c_k\binom{N}{k}^{1/2}z^k
\eeq
and the coefficients $c_k$ can be obtained from the Majorana polynomials as follows:
\[
c_k =  \frac{N+1}{\pi}\int_{\mathbb{C}} P_{N}(z)\binom{N}{k}^{1/2}(z^*)^k \left(1+\abs{z}^2\right)^{-(N+2)} d^2z.
\]
Some identities will be useful.
\subsubsection{Incomplete gamma function}
The following relations hold~\cite{abramowitz_64} 
\begin{align}
	\label{eq:gammaidentities}
	2\int_0^x r^{2k+1}e^{-r^2}dr & = \int_0^{x^2} u^k e^{-u} du \\
	\frac{1}{k!}\int_{0}^x u^k e^{-u}du	&= 1 -\left(1+x+\frac{x^2}{2!} + \frac{x^3}{3!} + \cdots +  \frac{x^k}{k!} \right) e^{-x}
\end{align}
\subsubsection{Majoration of the truncated exponential sum}
For all positive real $x$  and integer $k\geq 2$ such that $\frac{x}{k} >1$, we have the following upper bound 
\beq
\label{eq:majorationTruncExp}
\sum_{m=0}^{k} \frac{x^m}{m!} \leq \frac{x^k}{k!}
\left[
1 + \frac{1}{\frac{x}{k} -1}
\right].
\eeq
It  can be obtained as follows:
\[
\sum_{m=0}^{k} \frac{x^m}{m!}
= \left(\frac{x^k}{k!}\right)\left[
1 +  \sum_{m=0}^{k-1} k(k-1)\cdots (m+1) x^{m-k}
\right]
\]
but $k(k-1)\cdots (m+1) \leq k^{k-m}$ hence
\[
\sum_{m=0}^{k} \frac{x^m}{m!} \leq
\frac{x^k}{k!}\left[
1 +  \sum_{m=0}^{k-1} \left(\frac{x}{k}\right)^{m-k}
\right] = 
\frac{x^k}{k!}\left[
1 +  \left(\frac{k}{x}\right)^k\sum_{m=0}^{k-1} \left(\frac{x}{k}\right)^{m}
\right]
\]
the geometric sum can be calculated explicitly:
\[
\sum_{m=0}^{k-1} \left(\frac{x}{k}\right)^{m} = \frac{\left(\frac{x}{k}\right)^k-1}{\frac{x}{k} -1} \leq \frac{\left(\frac{x}{k}\right)^k}{\frac{x}{k} -1}
\]
where we have supposed that $x > k$.
Finally, we obtain
\[
\sum_{m=0}^{k} \frac{x^m}{m!} \leq \frac{x^k}{k!}
\left[
1 + \frac{1}{\frac{x}{k} -1}
\right].
\]
\subsubsection{Binomial coefficients}
We have the following results
\begin{align}
	\label{eq:binomargt}
	\binom{N}{k}\frac{k!}{N^k} &= \left(1-\frac{1}{N}\right)\left(1-\frac{2}{N}\right)\left(1-\frac{3}{N}\right)\cdots \left(1-\frac{k-1}{N}\right) \leq 1; \; 0\leq k \leq N  \\
	\label{eq:BinomApprox}
	& = 1 - \frac{k(k-1)}{2N} + \mathcal{O}(\frac{k^3}{N^2}).\\
	\label{eq:BinomJoseMaj}
	\ln \left( \binom{N}{k} \frac{k!}{N^k} \right) &\leq N [ -1 + x(1 - \ln ( x )) ]; \forall k\geq 2
\end{align}
where we set $x = 1 - (k-1)/N$ to simplify notations.
The upper bound of the logarithm can be obtained as follows.
Take the logarithm of Eq.~\eqref{eq:binomargt}
and use the fact that $t \mapsto \ln(1-t)$ is decreasing, then :
\begin{align*}
	\ln \left( \binom{N}{k} \frac{k!}{N^{k}} \right) &= \sum_{j=1}^{k-1} \ln \left( 1 - \frac{j}{N} \right)\\
	&\leq \int_{0}^{k-1} \ln \left( 1 - \frac{t}{N} \right) dt\\
	&= N \int_{1 - (k-1)/N}^1 \ln(u) du\\
	&= N [u\ln u - u]_{x}^1\\
	&= N(-1 - x\ln(x) + x)\\
	&= N(-1 + x(1 - \ln(x))).
\end{align*}

\section{Necessary condition for normalization}

We will show in this section the reciprocal of what was shown in the main article: If a Majorana polynomial is normalized with a Gaussian measure, then it can always be arbitrarily well approached by a truncated polynomial of degree $K \ll \sqrt{N}$ on a given region. For this we define the following integral
\beq
\label{eq:INR}
I(N,R) =  \frac{1}{\pi}\int_{\abs{z}\leq R}\abs{P_N\left(\frac{z}{\sqrt{N}}\right)}^2e^{-\abs{z}^2} dz_rdz_i
\eeq
where 
$P_N(z)$ is the Majorana polynomial of order $N$, given by Eq.~\eqref{eq:majorPoly},    associated to some state specified by its coefficients $c_k$ ($k=0,1,\cdots, N$), satisfying $\sum_{k=0}^N\abs{c_k}^2=1$.

\subsection{Part 1: Controlling the coefficients}

Let $I_{\mathbb{C}}(N) \equiv \lim_{R \rightarrow \infty} I(N,R)$.
We'll show that if $\lim_{N\rightarrow \infty} I_{\mathbb{C}}(N) = 1$, then we get the following
\be\label{sta} \forall \varepsilon > 0, \forall \delta > 0, \exists N_0, \forall N \geq N_0, \exists K < \delta\sqrt{N}, \sum_{k=K+1}^N |c_{k,N}|^2 \leq \varepsilon, \ee
where we indexed the coefficients of $P_N$ as $c_{k,N}$. Statement \eqref{sta} indeed corresponds to what is claimed in the main text: $K \ll \sqrt{N}$.\newline
Fix $\varepsilon > 0$, we will explicitly construct these $K$. For all $N \geq 0$ let
$K(N)$ be the smallest index such that
\[ \sum_{k=K(N)+1}^N |c_{k,N}|^2 \leq \varepsilon. \]
Being the smallest, we also get
\[ \sum_{k=K(N)}^N |c_{k,N}|^2 > \varepsilon. \]

We now show that for all $\delta > 0$, there exists $N_0$ such that for all $N \geq N_0$,
$K(N) < \delta \sqrt{N}$.\newline
We will show that by contradiction. So, we assume that for a certain $\delta > 0$ and for all $N_0$, there exists $N \geq N_0$
such that $K(N) \geq \delta \sqrt{N}$, and we would like to show that in this case $I_{\mathbb{C}}(N)$ can not converge to 1.
Note that $\delta$ is a constant independant of $N$.\newline
To simplify the notation, in the following we replace $K(N)$ with $K$. We then write :
\begin{align*}
    I_{\mathbb{C}}(N) &= \sum_{k=0}^N |c_{k,N}|^2 \binom{N}{k}\frac{k!}{N^k}\\
    &= \sum_{k=0}^{K-1} |c_{k,N}|^2 \binom{N}{k}\frac{k!}{N^k} + \sum_{k=K}^N |c_{k,N}|^2 \binom{N}{k}\frac{k!}{N^k}\\
    &= H(K, N) + T(K, N).
\end{align*}

Observe that the sequence $\binom{N}{k}\frac{k!}{N^k}$, indexed by $k$, is decreasing as it can
be seen in \eqref{eq:binomargt}. From that, we deduce :
\[ H(K, N) \leq \sum_{k=0}^{K-1} |c_{k,N}|^2 \binom{N}{0} \frac{0!}{N^0} = \sum_{k=0}^{K-1} |c_{k,N}|^2.\]
We also get, using again the fact that the sequence is decreasing :
\[ T(K, N) \leq \sum_{k=K}^N |c_{k,N}|^2 \binom{N}{K} \frac{K!}{N^{K}}. \] 
All that is left is to appropriately upper bound $\binom{N}{K} \frac{K!}{N^{K}}$. For this we use Eq.~\eqref{eq:BinomJoseMaj},
with $x = 1 - (K-1)/N$:
\[
    \ln \left( \binom{N}{K} \frac{K!}{N^K} \right) \leq N [ -1 + x(1 - \ln ( x )) ]; \forall K\geq 2.
\]
Now set $N$ to be large enough to have $x \approx 1 - \frac{K}{N}$ and $\frac{\delta}{\sqrt{N}} \ll 1$.
Our assumption that $K\geq \delta\sqrt{N}$ gives $x \leq 1 - \frac{\delta}{\sqrt{N}}$. If we now observe that $x(1-\ln(x))$ is increasing
for $x \leq 1$ 
\begin{align*}
    \ln \left( \binom{N}{K} \frac{K!}{N^{K}} \right)  
    &\leq N \left[ -1 + \left( 1 - \frac{\delta}{\sqrt{N}} \right) \left( 1 - \ln \left(1 - \frac{\delta}{\sqrt{N}} \right) \right) \right]\\
    &= N \left[ -1 + \left( 1 - \frac{\delta}{\sqrt{N}} \right) \left( 1 + \frac{\delta}{\sqrt{N}} +\frac{\delta^2}{2N} +  \mathcal{O}\left(\frac{1}{N^{3/2}}\right)\right) \right]\\
    &= -\frac{\delta^2}{2} + \mathcal{O}\left(\frac{1}{N^{1/2}}\right)
\end{align*}
Going back to exponential, we get the following bound:
\[ \binom{N}{K} \frac{K!}{N^K} \leq e^{-\delta^2/2}\left[1+ \mathcal{O}\left(\frac{1}{N^{1/2}}\right)\right] < 1
,\]
where the r.h.s. brackets can be made arbitrarily close to one since $N_o$ can be made arbitrarily large. It follows:
\begin{align*}
    I_{\mathbb{C}}(N) &= H(K, N) + T(K, N)\\
    &\leq \sum_{k=0}^{K-1} |c_{k,N}|^2 + e^{-\delta^2/2} \sum_{k=K}^N |c_{k,N}|^2\\
    &\leq\sum_{k=0}^{K-1} |c_{k,N}|^2 + \sum_{k=K}^N |c_{k,N}|^2 + (e^{-\delta^2/2}-1) \sum_{k=K}^N |c_{k,N}|^2\\
    &= 1 + (e^{-\delta^2/2}-1) \sum_{k=K}^N |c_{k,N}|^2\\
    &< 1 + (e^{-\delta^2/2}-1) \varepsilon < 1.
\end{align*}
So $I_{\mathbb{C}}(N)$ cannot converge to $1$, because it is bounded above by $1 + (e^{-\delta^2}-1) \varepsilon$
for infinitely many $N$. We thus end with a contradiction, as $P_N$ was assumed normalized. So $K(N)$ indeed satisfies :
\[ \forall \delta > 0, \exists N_0, \forall N \geq N_0, K(N) \leq \delta \sqrt{N}. \]
Informally, we showed that
\[ \forall \varepsilon > 0, \exists N_0, \forall N \geq N_0, \exists K \ll \sqrt{N}, \sum_{k=K+1}^N |c_{k,N}|^2 \leq \varepsilon. \]

\subsection{Part 2: Truncating the polynomial}
We will show that for $N$, and $K \ll N$ such that
\[ \sum_{k=K}^N |c_k|^2 \leq \varepsilon, \]
there exists $0 < R' < \sqrt{N}$ such that for $|z| \leq R'$
\[ \left| P_N \left( \frac{z}{\sqrt{N}} \right) - P_{N,K} \left( \frac{z}{\sqrt{N}} \right) \right|^2 \leq \varepsilon. \]
Notice that $R'$ is not necessarily the same as $R$, used in the integral \eqref{eq:INR}. We first use Cauchy-Schwarz for $|z| \leq R'$ where $R'$ is any positive radius for now :
\begin{align*}
    \left| P_N \left( \frac{z}{\sqrt{N}} \right) - P_{N,K} \left( \frac{z}{\sqrt{N}} \right) \right|^2
    &= \left| \sum_{k=K}^N c_k \binom{N}{k}^{1/2} \frac{z^k}{N^{k/2}} \right|^2\\
    &\leq \sum_{k=K}^N |c_k|^2 \sum_{k=K}^N \binom{N}{k} \frac{|z|^{2k}}{N^{k}}\\
    &\leq \varepsilon \sum_{k=K}^N \binom{N}{k} \frac{R'^{2k}}{N^{k}}.
\end{align*}
If we then make the assumptions that $R'^2 \leq K$ and that $R'^{2K} \leq K!$ (when $K\gg1$ these two conditions are nearly equivalent). Hence for all $k \geq K$, we have
$R'^{2k} \leq k!$ and we can upper bound the difference between the polynomials by the following sum that we have already studied :
\begin{align*}
    \left| P_N \left( \frac{z}{\sqrt{N}} \right) - P_{N,K} \left( \frac{z}{\sqrt{N}} \right) \right|^2
    &\leq \varepsilon \sum_{k=K}^N \binom{N}{k} \frac{k!}{N^{k}}\\
    &\leq \varepsilon N \binom{N}{K} \frac{K!}{N^K}.
\end{align*}
Upper bounding the log of $\binom{N}{K} \frac{K!}{N^K}$ as before gives:
\begin{align*}
    \ln \left( \binom{N}{K} \frac{K!}{N^K} \right) &\leq N \left[-1 - \left( 1 - \frac{K}{N} \right) \ln \left( 1 - \frac{K}{N} \right) + 1 - \frac{K}{N} \right]\\
    &= N \left[ \left( 1 - \frac{K}{N} \right) \left( - \frac{K}{N} + O\left( \frac{K^2}{N^2} \right) \right) - \frac{K}{N} \right]\\
    &= -2K + {\mathcal O} \left( \frac{K^2}{N} \right)
\end{align*}
Assume we have $\frac{\ln(N)}{2} \leq K$. This can always be achieved, if $\sum_{k=K}^N |c_k|^2 \leq \varepsilon$,
and this is also true for all $K' \geq K$. Then
\begin{align*}
    \ln \left( \binom{N}{K} \frac{K!}{N^K} \right) &\leq - \ln(N) + o(1)
\end{align*}
If we then go back to exponential, we get
\[ \binom{N}{K} \frac{K!}{N^K} \leq \frac{1}{N} \]
By putting all of these together, we achieve
\begin{align*}
    \left| P_N \left( \frac{z}{\sqrt{N}} \right) - P_{N,K} \left( \frac{z}{\sqrt{N}} \right) \right|^2
    &\leq \varepsilon N \frac{1}{N}\\
    &\leq \varepsilon
\end{align*}
for all $|z| \leq R'$.

\section{Convergence of the integral}

We will show that 

\be 
\text{ if } \forall \epsilon>0 \; \exists K_0 \in \mathbb{N}, \text{ s.t } \forall K\geq K_0 \; \exists N > K  \text{ s.t } \sum_{k=K+1}^N \abs{c_k}^2\leq \epsilon,
\ee
$\text{then } \exists R \text{ s.t. }   \abs{I(N,R) -1} \le 4\epsilon$; where $I(N,R)$ is given by Eq.~\eqref{eq:INR}.
	
Inserting this expression of  $P_N(z)= \sum_{k=0}^N c_k \binom{N}{k}^{1/2} z^k$ in 
the integral $I(N,R)$ and calculating the integral in polar coordinates  $z=re^{i\theta}$ we obtain
\[
I(N,R) = \sum_{k=0}^{N} \abs{c_k}^2 \binom{N}{k}\frac{2}{N^k} \int_{0}^{R} 
r^{2k+1} e^{-r^2} dr.
\]
We can obtain an explicit expression of the integral $I(N,R)$ using the identities given by Eqs.\eqref{eq:gammaidentities};
we obtain,
\[
I(N,R) = \sum_{k=0}^N \abs{c_k}^2\binom{N}{k}\frac{k!}{N^k} 
\left[1 -\left(1+R^2+\frac{R^4}{2!} +  \cdots +  \frac{R^{2k}}{k!} \right) e^{-R^2} \right]
\]
where as before (see Eq.~\eqref{eq:binomargt}) $\binom{N}{k}\frac{k!}{N^k}$ can be written as
\[
\binom{N}{k}\frac{k!}{N^k} = \left(1-\frac{1}{N}\right)\left(1-\frac{2}{N}\right)\left(1-\frac{3}{N}\right)\cdots \left(1-\frac{k-1}{N}\right) <1.
\]

Now we can make assumptions on the sequence $c_k$ an see explicitly  their effect in the integral. Suppose that for an arbitrary small positive $\epsilon$ we can find two integers  $K_0$ and $N$ such that for all $K\ge K_0$ we have
$\sum_{k=K+1}^N \abs{c_k}^2\leq \epsilon$.
Then the sum representing the integral can be divided in 2 terms:
\[
I(N,R) = S(K,N,R) + \overline{S}(K,N,R)
\]
where 
\[
S(K,N,R) = \sum_{k=0}^K \abs{c_k}^2\binom{N}{k}\frac{k!}{N^k} \left[1 -\left(1+R^2+\frac{R^4}{2!} +  \cdots +  \frac{R^{2k}}{k!} \right) e^{-R^2} \right]
\]
and 
\[
\overline{S}(K,N,R) =  \sum_{k=K+1}^N \abs{c_k}^2\binom{N}{k}\frac{k!}{N^k} \left[1 -\left(1+R^2+\frac{R^4}{2!} +  \cdots +  \frac{R^{2k}}{k!} \right) e^{-R^2} \right]
\]
It it is clear that $\overline{S}(K,N,R)  \leq \epsilon$.
Indeed the factor in the square bracket is less that 1 and the factor $\binom{N}{k}\frac{k!}{N^k}$ is also less than 1 because of Eq.~\eqref{eq:binomargt};
hence
\[
\overline{S}(K,N,R) < \sum_{k=K+1}^N \abs{c_k}^2 \leq \epsilon.
\]
We thus conclude that
\[
I(N,R) - S(K,N,R) = \overline{S}(K,N,R) \leq \epsilon.
\]
We note that this inequality is completely independent of $R$.

Now we can find conditions for $K$ and $R$ such that $S(K,N,R)$ is as close to 1 as we wish.
For that we write 
$S(K,N,R) = A(K,N) - \overline{A}(K,N,R)$
where
\[
A(K,N) = \sum_{k=0}^K \abs{c_k}^2\binom{N}{k}\frac{k!}{N^k} = 
\sum_{k=0}^K \abs{c_k}^2 \left(1-\frac{1}{N}\right)\left(1-\frac{2}{N}\right)\left(1-\frac{3}{N}\right)\cdots \left(1-\frac{k-1}{N}\right) 
\]
and 
\[
\overline{A}(K,N,R) = \sum_{k=0}^K \abs{c_k}^2\binom{N}{k}\frac{k!}{N^k} 
\left(1+R^2+\frac{R^4}{2!} +  \cdots +  \frac{R^{2k}}{k!} \right) e^{-R^2}
\]
To obtain an upper bound of $\overline{A}(K,N,R)$ we can use the upper bound of Eq.~\eqref{eq:majorationTruncExp}.
Using this relation we obtain
\[
\overline{A}(K,N,R) \leq 
\sum_{k=0}^{K} \abs{c_k}^2 \frac{R^{2k}}{k!}e^{-R^2}
\left[
1 + \frac{1}{\frac{R^2}{k}-1}
\right]
\]
assuming that $\sum_{k=0}^K \abs{c_k}^2\leq 1$ we obtain
\[
\overline{A}(K,N,R)\leq  R^{2K}e^{-R^2}
\left[
1 + \frac{1}{\frac{R^2}{K}-1}
\right].
\]
To have  $\overline{A}(K,N,R)\leq \epsilon$ it is thus sufficient that
\[
R^{2K}e^{-R^2}
\left[
1 + \frac{1}{\frac{R^2}{K}-1}\right]
\leq \epsilon
\] 
which can be written as
\[
R^2 - K\ln(R^2) -\ln(1 + \frac{1}{\frac{R^2}{K}-1}) \geq \ln{\left (\frac{1}{\epsilon}\right )}
\] 
This allows to write
\[
I(N,R) - A(K,N) \leq 2\epsilon.
\]
Using Eq.~\eqref{eq:binomargt} we can write
\begin{align*}
	A(K,N) &=
	\sum_{k=0}^K \abs{c_k}^2 \left(1-\frac{1}{N}\right)\left(1-\frac{2}{N}\right)\left(1-\frac{3}{N}\right)\cdots \left(1-\frac{k-1}{N}\right) \\
	&\simeq
	\sum_{k=0}^K \abs{c_k}^2 \left[1 - \frac{k(k-1)}{2N}\right] = 1-\epsilon - \sum_{k=0}^K \abs{c_k}^2\frac{k(k-1)}{2N}
\end{align*}
and the last term can be made smaller than $\epsilon$ choosing $N$ such that
\[
\frac{K(K-1)}{2N} \leq \epsilon.
\]

\section{Uniform convergence between Majorana polynomial and Bargmann function}

In this section, we explicitly prove the uniform convergence of the Majorana polynomial to the Bargmann function when the CV limit is assumed, {\it i.e.}, for $\abs{z}^2 \ll \sqrt{N}$. We start by recalling that a sequence of function $(f_n)_n$ is said to converge uniformly towards a function $f$ on a domain $D$ if: 
\begin{equation}
    \forall \epsilon>0, \exists N \in \mathbb{N}, \forall z\in I, \forall n \geq N, |f_n(z)-f(z)|\leq \epsilon.
\end{equation}

\subsection{Uniform convergence for truncated states }
\label{sec:UniformCov}
Consider the family of SSRC states $\ket{\psi^N}_{S } =\sum_{k=0}^{K} c_k \ket{k}_a\ket{N-k}_b$, $K<N$, with the associated Majorana polynomials $P_N(z)$. Let's also consider the CV state  $\ket{\psi}_{C} = \sum_{k=0}^{K} c_k \ket{k}$ and its associated Bargmann function $B(z)$. The domain $D=\{ z\in \mathbb{C}; |z|\leq R\}$. 
Let's prove the uniform convergence of the family of polynomials $(P_N)_{N>K}\left(\frac{z}{\sqrt{N}}\right)$ towards $B(z)$ for $z\in D$.   
\begin{proof}
\begin{align*}
   \left|P_N\left(\frac{z}{\sqrt{N}}\right)-B\left(z\right)\right|^2 &=\left| \sum_{k=0}^K \frac{c_kz^k}{\sqrt{k!}}\left(1-\left(\frac{\binom{N}{k}k!}{N^k} \right)^{1/2}\right)\right|^2\\
   &\leq (K+1) \sum_{k=0}^K \frac{|c_kz^k|^2}{{k!}}\left|1-\left(\frac{\binom{N}{k}k!}{N^k} \right)^{1/2}\right|^2\\
\end{align*}

Using the Stirling approximation (see Eq.\eqref{eq:BinomApprox}), we have that $ \left|1-\left(\frac{\binom{N}{k}k!}{N^k} \right)^{1/2}\right|=  \frac{k(k-1)}{2N} + \mathcal{O}(\frac{k^3}{N^2})$. Consequently,  for $N \gg  K^2$,  $\mathcal{O}(\frac{k^3}{N^2})\leq \frac{k(k-1)}{2N}$, hence $\left|1-\left(\frac{\binom{N}{k}k!}{N^k} \right)^{1/2}\right|^2 \leq  (\frac{K^2}{N})^2$ for $k\leq K$. 

Using this gives:

\begin{align*}
     \left|P_N\left(\frac{z}{\sqrt{N}}\right)-B\left(z\right)\right|^2 &\leq (K+1) \left(\frac{K^2}{N}\right)^2 \sum_{k=0}^K \frac{|c_kz^k|^2}{{k!}}\\
   &\leq (K+1) \left(\frac{K^2}{N}\right)^2 \sum_{k=0}^K \frac{R^{2k}}{{k!}}\\
   & \leq (K+1) \left(\frac{K^2}{N}\right)^2 \frac{R^{2K}}{K!}
   \left[
   1 + \frac{1}{1-\frac{R^2}{K}}.
   \right]
\end{align*}
where in the last inequality we have used Eq.~\eqref{eq:majorationTruncExp}.

We see that choosing $K$ such that $K^2\leq \epsilon N$ as before and $R$ such that
\[
 \frac{R^{2K}}{K!} (K+1)
\left[
1 + \frac{1}{1-\frac{R^2}{K}}
\right] \leq
1
\] 
it is sufficient to ensure that  $ \left|P_N\left(\frac{z}{\sqrt{N}}\right)-B\left(z\right)\right|^2\leq \epsilon^2$. 
For sufficient large value of $K$ the condition on $R$ amounts to $R^2\leq K$. Indeed taking the logarithm  the condition becomes
\[
K\ln(R^2) \leq \ln(K!) -\ln\left((K+1)
\left[
1 + \frac{1}{1-\frac{R^2}{K}}
\right]  \right)
\] 
and using the Stirling approximation  $\ln(K!)\simeq K\ln(K) - K$ we obtain
\[
\ln(R^2) \leq \ln(K) - 1 - 
\frac{1}{K}\ln\left((K+1)
\left[
1 + \frac{1}{1-\frac{R^2}{K}}
\right]  \right)
 \simeq \ln(K).
\]

\end{proof}

\subsection{Uniform convergence for Majorana states satisfying the CV limit}

Consider the family of states $\ket{\psi^N}_{S }$ parametrized by the total number of photons $N$ :    $\ket{\psi^N}_{S } =\sum_{k=0}^{N} c_{k, N} \ket{k}_a\ket{N-k}_b$ with $\sum_{k=0}^{N} | c_{k, N}|^2 =1$. Additionally, suppose the CV limit for this family of states, that is: 
\begin{eqnarray}
    \forall \epsilon>0, \exists N\in \mathbb{N} / \forall M \geq N, \exists K << M, \sum_{k=K}^M |c_{k,M}|^2\leq \epsilon
\end{eqnarray}

Call $(P_{N})_N$ the family of Majorana polynomial, and $P_{K,N}$ the truncated ones. Let's also extends those coefficients by defining for $k>N,  c_{k, N} = 0$. Let's call $c_k$ the limit of the coefficients $c_{k, N}$, and assume the convergence in the following sense:
\begin{eqnarray}
    \forall \epsilon>0, \exists N\in \mathbb{N}/ \forall M\geq N,\sum_{k=0}^\infty |c_{k,M}-c_k|\leq \epsilon
\end{eqnarray} 

Use this infinite set of coefficients $c_k$ to define the CV state $\ket{\psi}_{C} = \sum_{k=0}^{\infty} c_k \ket{k}$,  $ \sum_{k=0}^\infty|c_{k}|^2 =1 $. Call $B$ its associated Bargmann function. The aim is to prove the uniform convergence of the family of Majorana polynomial $(P_{N})_N$ towards the Bargmann function $B$. 

\begin{proof}
Choose $\epsilon >0$.
By hypothesis, there exists $N_1/\forall M>N_1, \exists K<<M / \sum_{k=K}^{M} |c_{k,M} |^2 \leq \epsilon^2$. Hence from part II.B, there exists $0 < R< \sqrt{N}$ such that for $|z| \leq R$, $\left| P_M \left( \frac{z}{\sqrt{M}} \right) - P_{M,K} \left( \frac{z}{\sqrt{M}} \right) \right| \leq \varepsilon.$

Using the second hypothesis, we also have that: $\exists N_2\in \mathbb{N}/ \forall M\geq N_2,\sum_{k=0}^\infty |c_{k,M}-c_k|\leq \frac{\epsilon}{R^k}$. Remark that it gives both $\sum_{k=0}^M |c_{k,M}-c_k|\leq \frac{\epsilon}{R^k}$ and $\sum_{k=M+1}^\infty |c_k|\leq \frac{\epsilon}{R^k}$.

Take $N=\max(N_1, N_2)$, for $z/ |z|\leq R$ it follows that for $n\geq N$, 

\begin{align*}
 \left|P_n\left(\frac{z}{\sqrt{n}}\right)-B\left(z\right)\right| &=  \left|P_n\left(\frac{z}{\sqrt{n}}\right)-P_{K,n}\left(\frac{z}{\sqrt{n}}\right)  + P_{K,n}\left(\frac{z}{\sqrt{n}}\right)-B\left(z\right)\right|\\
    &\leq \left|P_n\left(\frac{z}{\sqrt{n}}\right)-P_{K,n}\left(\frac{z}{\sqrt{n}}\right)\right| + \left|P_{K,n}\left(\frac{z}{\sqrt{n}}\right)-B\left(z\right)\right|
\end{align*}

We then have the following bounds:
\begin{itemize}
    \item \(|P_n\left(\frac{z}{\sqrt{n}}\right)-P_{K,n}\left(\frac{z}{\sqrt{n}}\right)|\leq \epsilon\) due to Part II.B.
    \item For \(|P_{K,n}\left(\frac{z}{\sqrt{n}}\right)-B\left(z\right)|\), decompose \(B\left(z\right)\) into two parts, \(B_K\left(z\right) =   \sum_{k=0}^K \frac{c_kz^k}{\sqrt{k!}}\) and \( \sum_{k=K+1}^\infty \frac{c_kz^k}{\sqrt{k!}}\). Hence,  \(|P_{K,n}\left(\frac{z}{\sqrt{n}}\right)-B\left(z\right)|  \leq |P_{K,n}\left(\frac{z}{\sqrt{n}}\right)-B_K\left(z\right)| + | \sum_{k=K+1}^\infty \frac{c_kz^k}{\sqrt{k!}}| \). Both of those terms can be made infinitesimal thanks to the condition of the CV limit:
    \begin{itemize}
        \item Introduce the function \(\Tilde{B}_K(z) =\sum_{k=0}^K\frac{c_{k, n} z^k}{\sqrt{k!}} \).   Then, \(|P_{K,n}\left(\frac{z}{\sqrt{n}}\right)-B_K\left(z\right)| \leq |P_{K,n}\left(\frac{z}{\sqrt{n}}\right)-\Tilde{B}_K(z)| + |\Tilde{B}_K(z) -B_K\left(z\right)| \).
        Firstly, from  part \ref{sec:UniformCov}  \(|P_{K,n}\left(\frac{z}{\sqrt{n}}\right)-\Tilde{B}_K(z)| \leq \sqrt{ (K+1)} \frac{K^2}{n}\frac{R^{K}}{\sqrt{K!}}
   \sqrt{1 + \frac{1}{1-\frac{R^2}{K}}}\)  goes to zero for \(|z|\leq R\). Secondly,  \(|\Tilde{B}_K(z) -B_K\left(z\right)| \leq \sum_{k=0}^K |c_{k,n}-c_k|\cdot \frac{R^k}{\sqrt{k!}}\leq \epsilon \) by hypothesis.
        \item Secondly, let's use Cauchy-Schwartz inequality to bound \(\left| \sum_{k=K+1}^\infty \frac{c_kz^k}{\sqrt{k!}}\right|\):
		\begin{align*}
		     \left| \sum_{k=K+1}^\infty \frac{c_kz^k}{\sqrt{k!}}\right|&\leq \sqrt{ \sum_{k=K+1}^\infty  \left| \frac{z^k}{\sqrt{k!}} \right|^2  \cdot \sum_{k=K+1}^\infty \left|c_k\right|^2}\\
             &= \sqrt{\sum_{k=K+1}^\infty   \frac{\left|z\right|^{2k}}{k!} }   \cdot \sqrt{\sum_{k=K+1}^\infty \left|c_k\right|^2}\\
		\end{align*}
        First term is the rest of the exponential serie, and can be smaller than $\epsilon$ for sufficiently large $K$ at fixed $R$, while second term is also infinitesimal due to the two hypothesis. 
	\end{itemize}
\end{itemize}

Altogether, the three terms appearing in there can be made arbitrarily small, hence proving the uniform convergence for states in the CV limit.
\end{proof}

\subsection{Convergence of Majorana zeros towards Bargmann zeros}
In this section, we study the convergence of the Majorana polynomial to the Bargmann function in the CV limit, through the zeros of the respective functions. For a fixed $n <N$, consider $\ket{\psi}_{S } =\sum_{k=0}^{n} c_k \ket{k}_a\ket{N-k}_b$ and $\ket{\psi}_{C} = \sum_{k=0}^{n} c_k \ket{k}$. The SSRC (CV) state can be parametrized by the set of rotations $\{\alpha_k\}$ (displacements $\{\beta_k\}$) as follows: 
	\begin{equation}\label{s_eq}
		\ket{\psi}_{S} =\sum_{k=0}^{n} c_k \ket{k}_a\ket{N-k}_b= \mathcal{N}\prod_{k=1}^{n} R(\alpha_k,N)\hat{a}^\dagger R^\dagger(\alpha_k, N) \cdot (\hat{b}^\dagger)^{N-n} \ket{\emptyset}
	\end{equation}
	\begin{equation}\label{CV general}
		\ket{\psi}_{C} = \sum_{k=0}^{n} c_k \ket{k} = \frac{c_n}{\sqrt{n!}}\prod_{k=1}^{n} D(\beta_k)\hat{a}^\dagger D^\dagger(\beta_k)  \ket{\emptyset}
	\end{equation}
The set $\{\alpha_k\}$ is the set of zeros of the truncated Majorana polynomial $P_{n,N}\left(\frac{z}{\sqrt{N}}\right)$ associated to $\ket{\psi}_{S}$, $P_{n,N}\left(\frac{z}{\sqrt{N}}\right)= \sum_{k=0}^n\frac{c_k\binom{N}{k}^{1/2}}{N^{k/2}}z^k$. Similarly, the set $\{\beta_k\}$ is the set of zeros associated to the truncated Bargmann function $B_n(z)=\sum_{k=0}^n \frac{c_k}{\sqrt{k!}}z^k $. For $n \ll N$ the coefficients become asymptotically close implying an asymptotic equality between the sets $\{\alpha_k\}$ and $\{\beta_k\}$. 

First, consider the following example:  we will prove that zeros of Majorana polynomial of $c_0\ket{0}\ket{N} + c_1\ket{1}\ket{N-1}+ c_2\ket{2}\ket{N-2}$ become asymptotically close for $N\gg1$ to the zeros of the Bargmann function of $c_0\ket{0} + c_1\ket{1}+ c_2\ket{2}$.
\begin{proof}	
On the CV side, the state $\ket{\psi}_C = c_0\ket{0}+c_1\ket{1} +c_2\ket{2}$ can be rewritten as $\ket{\psi}_C =  \frac{c_2}{\sqrt{2}}(a^\dagger-\beta_1^*)(a^\dagger-\beta_0^*)\ket{\emptyset}$. Coefficients $\beta_{0,1}$ are related to the coefficients $c_k$ by identification of the coefficients along the Fock basis. It gives:
	\begin{equation}\label{identification}
		c_0  =\frac{c_2}{\sqrt{2}}\beta_1^*\beta_0^* \quad \text{and } \quad c_1=-\frac{c_2}{\sqrt{2}}(\beta_1^*+\beta_0^*)
	\end{equation}
	The two sets of solutions, $(\beta_{0,-},\beta_{1-}) $ and $(\beta_{0,+},\beta_{1+}) $, yield two states : $\ket{\psi_-}_C = \frac{c_2}{\sqrt{2}}(a^\dagger-\beta_{1-}^*)(a^\dagger-\beta_{0-}^*)\ket{\emptyset}$ and $\ket{\psi_+}_C = \frac{c_2}{\sqrt{2}}(a^\dagger-\beta_{1+}^*)(a^\dagger-\beta_{0+}^*)\ket{\emptyset}$, that are equal as one checks that $(\beta_{0-},\beta_{1-})  = (\beta_{1+},\beta_{0+}) $. The set of equations between $\{c_k\}$ and $\{\beta_k\}$ is non linear, yielding multiple solutions. However, those are simply permutations, translating the underlying commutation relation : $(a^\dagger-\beta_1^*)(a^\dagger-\beta_0^*) = (a^\dagger-\beta_0^*)(a^\dagger-\beta_1^*)$.\\
	
 On the SSRC side, state $\ket{\psi}_S = c_0\ket{0}\ket{N} + c_1\ket{1}\ket{N-1}+ c_2\ket{2}\ket{N-2}$ can be rewritten as 
 \be
 \ket{\psi}_S= \mathcal{N}^{-1} R(\alpha_0, N )\hat{a}^\dagger R^\dagger(\alpha_0, N ) R(\alpha_1, N )\hat{a}^\dagger R^\dagger(\alpha_1, N ) (\hat{b}^\dagger)^{N-2}\ket{\emptyset}.
 \ee 
Expanding and identifying the terms in the SSRC basis relates $\{c_k\}$ and $\{\alpha_k\}$. It gives
	\begin{align}
		c_0  &=\frac{c_2}{\sqrt{2}}\alpha_1^*\alpha_0^* \left(1-\frac{(|\alpha_0|^2+ |\alpha_1|^2+1)}{2N}+\mathcal{O}(\frac{1}{N^2}) \right)\\
        c_1&=-\frac{c_2}{\sqrt{2}}\left(\alpha_1^*+\alpha_0^* -\frac{\alpha_0^*(1+|\alpha_0|^2)+\alpha_1^*(1+|\alpha_1|^2)}{2N}+\mathcal{O}(\frac{1}{N^2})\right)
	\end{align}

which coincides with Eq. \ref{identification} in the limit $|\alpha_k|/N\ll1, 1\ll N $, hence yielding the same set of zeros. 
\end{proof}

Let's now treat the general case described above, and prove that asymptotically, the two sets of zeros $\{\alpha_k\}_k, \{\beta_k\}_k$ coincide.  

\begin{proof}

	\textbf{a)} Identifying both sides of Eq. \ref{CV general} along $\ket{k}$ yields the following relation between $\{\beta_k\}$ and $\{c_k\}$, where $\mathcal{P}_{n-k}([1,n])$ denote the set of all subset of $\{ 1, ...n \} $ of size $n-k$:
	
	\begin{align}\label{c_k function of beta_k}
		\forall k\in [0,n] ,  \frac{c_k}{\sqrt{k!}}=\frac{c_n}{\sqrt{n!}}(-1)^{n-k}\sum_{I\in\mathcal{P}_{n-k}([1,n]) }\prod_{i\in I}\beta^*_i 
	\end{align}
	\textbf{ b)} Identifying both sides of Eq. \ref{s_eq} along $\ket{k}_a\ket{N-k}_b$ yields the following relation between $\{\alpha_k\}$ and $\{c_k\}$:
    \begin{align*}
		\forall k\in [0,n],   \frac{c_k}{\sqrt{k!}}=\frac{c_n}{\sqrt{n!}}(-1)^{n-k}\sum_{I\in\mathcal{P}_{n-k}([1,n]) }\prod_{l\in I}\alpha^*_l\left(1-\frac{|\alpha_l|^2}{N}\right)^{-1/2}\frac{\sqrt{(N-k)!}}{\sqrt{(N-n)!}\sqrt{N}^{n-k}}
	\end{align*}
	
	In this set of equations, $\frac{\sqrt{(N-k)!}}{\sqrt{(N-n)!}\sqrt{N}^{n-k}} \sim 1 $ for $N\gg k,n$, and $\prod_{l\in I}\alpha^*_l\left(1-\frac{|\alpha_l|^2}{N}\right)^{-1/2}= \prod_{l\in I}\alpha^*_l + \mathcal{O}(1/N)$, assuming that $\alpha_l$ are of order 1, independent of N, i.e. that $\max_l(|\alpha_l|)\ll N $. Assuming this hypothesis fulfilled, one gets the simplified set of equations :
	\begin{align} \label{c_k function of alpha_k}
		\forall k\in [0, n],  \frac{c_k}{\sqrt{k!}}=\frac{c_n}{\sqrt{n!}}(-1)^{n-k}\sum_{I\in\mathcal{P}_{n-k}([1,n]) }\prod_{l\in I}\alpha^*_l + \mathcal{O}(1/N)
	\end{align}
	
	In \textbf{ b)}, the set of the $n$ parameters $\{\alpha_k\}$ is solution of the set of $n+1 $  equations given by Eqs. \ref{c_k function of alpha_k}. In \textbf{ a)},the set of the $n$ parameters $\{\beta_k\}$ is solution of the set of $n+1 $ equations given by Eqs. \ref{c_k function of beta_k}. Neglecting the $\mathcal{O}(1/N)$, Eqs. \ref{c_k function of alpha_k}  and \ref{c_k function of beta_k} form the same set of non linear-equations, and the parameters $\{\alpha_k\}$  and  $\{\beta_k\}$ are two solutions.  A priori, it is possible that  those are different solutions, ie $\{\alpha_k\} \neq \{\beta_k\}$. However, a state is described uniquely by the unordered set of zeros. The set of equations must yield one solution plus all possible permutation of it. This implies that the sets of parameters  $\{\alpha_k\}$  and  $\{\beta_k\}$ coincides well, $\{\alpha_k\}=\{\beta_k\}$. 
\end{proof}

\bibliography{BibStellar.bib}
\end{document}